\documentclass[12pt]{article}
\newcommand{\beq}{\begin{equation}}
\newcommand{\eeq}{\end{equation}}
\newcommand{\bea}{\begin{eqnarray}}
\newcommand{\eea}{\end{eqnarray}}
\begin{document}
\thispagestyle{empty}
\vspace*{-15mm}
%----------
\baselineskip 10pt
\begin{flushright}
\begin{tabular}{l}
%{\bf OCHA-PP-186}\\
{\bf SEPTEMBER 2009}\\
%{\bf hep-th/YYMMDD}
\end{tabular}
\end{flushright}
\baselineskip 24pt 
\vglue 10mm 
%%%%%%%%%%%%%%%%%%%%%%%%%%%%%%%%%%%%%%%%%%%%%% 
%                Title 
%%%%%%%%%%%%%%%%%%%%%%%%%%%%%%%%%%%%%%%%%%%%%%
\begin{center}
{\LARGE\bf{Three dimensional gravity and Quantum Hall Effect
}}
\vspace{7mm}\\
\baselineskip 18pt 
{\bf
Yuko KOBASHI}\\
\vspace{2mm}
{\it 
 Department of Physics, Ochanomizu University, \\
 2-1-1, Otsuka, Bunkyo-ku, Tokyo 112-8610, Japan
}\\
\vspace{10mm}
\end{center}
%%%%%%%%%%%%%%%%%%%%%%%%%%%%%%%%%%%%%%%%
%%%%%                              %%%%%
%%%%%          Abstract            %%%%%
%%%%%                              %%%%%
%%%%%%%%%%%%%%%%%%%%%%%%%%%%%%%%%%%%%%%%
\begin{center}
{\bf Abstract}\\[7mm]
\begin{minipage}{14cm}
\baselineskip 16pt
\noindent
%%%%%----------------------------------
Recently three dimensional Einstein gravity with AdS geometry has been studied, and pointed out to be described with Chern-Simons theory by Grumiller and Jackiw.  While, non-commutative Chern-Simons theory is known to be equivalent to Quantum Hall Effect, through Area Preserving Diffeomorphism by Susskind and others.  Conbining these, three dimensional gravity is studied in this paper in the context of the Quantum Hall Effect, and non-commutative three dimensional gravity is shown to be equivalent to the matrix formulated exciton model which is the two component matrix model previously by the author.
%%%%%----------------------------------  
\end{minipage}
\end{center}
%%%----------------------------------  
%%%
%%%
%%%   Main body of the paper
%%%
%%%
%%%----------------------------------  
\newpage
\baselineskip 18pt 
\def\thefootnote{\fnsymbol{footnote}}
\setcounter{footnote}{0}
%--------------------------------
%
%         Introduction
%
%--------------------------------
\section{Introduction}
Chern-Simons terms were introduced into physics, and have played various roles both in physical and mathematical contexts.  
Susskind~\cite{Susskind} has proposed a matrix model of the Chern-Simons type, and shown the equivalence of the model to quantum Hall effect.

The quantum Hall effect is described by the system of electrons flowing in 2-dim space, with the background strong uniform magnetic field.  Each electron is doing the cyclotron motion by the existence of the magnetic field, which gives rise to non-commutativity through the area that each electron sweeps and other electrons are excluded from this area.  That is, if we assume a pair of continuous coordinates $x_1, x_2$ which label the material points moving with the fluid, they gives the commutation relation $[x_a, x_b] = i\theta\epsilon_{ab}$, with $a,b = 1,2$.  Here $2\pi\theta$ is the excluding area of an electron from the others.  Namely, it is the inverse of the number density of $\rho$ of the electron,
%---
\begin{eqnarray}
 2\pi \theta = \rho^{-1}.
\end{eqnarray}
%---
Assume a pair of coordinates $y_1,y_2$, which are the labeling of the electrons, and the coordinates of the fluid can be described as $x_i(y,t)$ with $i=1,2$.  If the fluid field $x$ transform as a scalar, $x^{\prime}(y^{\prime})=x(y)$, and moreover, the coordinates $x$ and $y$ keep the Jacobian $|\frac{\partial y}{\partial x}| = 1$, the transformation is Area Preserving Diffeomorphisms (APD).  Any area is kept constant under APD, which means that the fluid is incompressive, that is the fluid has no vortices.  Then, assume the Jacobian is 1, where the real space density of the fluid is given as $\rho=\rho_0|\frac{\partial y}{\partial x}|$, the time independent equilibrium solution of the equation of motion is given by
%---
\begin{eqnarray}
 x_i=y_i.
\end{eqnarray}
%---
Now consider small deviation from this solution by introducing Chern-Simons gauge field $A$, and expand the original action of quantum Hall effect.  Then the action of the quantum Hall effect accords with the APD action~\cite{APD}, and what's more, it coincides with the action which can be gained through expanding star products of non-commutative Chern-Simons action to first order in $\theta$.  As a result, it is shown that the quantum Hall effect is equivalent to the non-commutative Chern-Simons theory through APD.

Then, one can gain the action like the matrix theory of D0-branes by replacing all the parameter of the action of quantum Hall effect.  To explain the way concretely, it is to replace the classical configuration space of K electrons by a space of two $K \times K$ hermitian matrices $X_a$ where $a=1,2$.
This Susskind's matrix-like model is reformulated by using finite matrix introducing an additonal boundary field.~\cite{finite-matrix}  
If one expresses particles in the fluid, one could do by using the analogy of vortices.  To construct the action of the quantum Hall effect, one should add constraint defining property of the fluid, with $A_0$, which is the time component of the Chern-Simons gauge field as the Lagrange constant to the action of 2-dim electron moving in background magnetic field.  If the fluid is incompressive, the constraint that is describing Jacobian is 1.  Then, one could include vortices to add $\delta$ function term to the constraint.  Here, vortices can be interpreted as surplus or deficit of electrons of the fluid, which is so called quasi-particle or quasi-hall respectively.  The area occupied by quasi-particle or quasi-hall proportions to the constant, called filling fraction, the ratio of the number of electrons to the magnetic flux,
%---
\begin{eqnarray}
 \nu\equiv \frac{2\pi\rho_0}{eB}
\end{eqnarray}
%---
which means an area occupied by a magnetic flux.  $1/\nu$ is the Chern-Simons level.  The delta function on the noncommutative space that is representing vortices should be replaced by a projection operator onto a particular vector in the matrix space~\cite{delta}.  Assume some basis vectors $|m)$ and $m$ runs over the positive integers to zero.  A delta function at the origin may be
%---
\begin{eqnarray}
 \theta\delta(y) \rightarrow |0)(0|.
\end{eqnarray}
%---   
Then, one would obtain the constraint to express quasi-particles;
%---
\begin{eqnarray}
 [x_1,x_2]=i\theta +i\theta\nu |0)(0|.
 \label{constraint-qe}
 \end{eqnarray}
%--- 
In Susskind's paper~\cite{Susskind}, the solution of the constraint equation for the quasi-electrons or the quasi-halls was already commented.  However, it's seem like to more natural when the model includes both quasi-electron and quasi-hall, and their bound state can make an exciton.  Then the model established by Susskind was extended to describe exciton by generalizing the constraint condition, and the equation of constraint was solved to obtain the solution of exciton in the infinite-sized matrix case by authors.  And an example of the physical states for the model was given also.  Moreover, more natural system was established by including Coulomb potential, and finally, the dispersion relation of exciton, of which behavior resembles magnetic roton was gained in the paper~\cite{our work}.

On the other hand, some effort was taken to quantize 3-dim Einstein gravity theory by many people.  One of the proposal was introducing Chern-Simons terms with coefficient $\frac{1}{\mu}$ in 3-dim Einstein action to make massive nature.  This is well known as topological massive gravity (TMG)~\cite{TMG}.  TMG has an $AdS_3$ solution.  It was shown in ~\cite{2D CFT} that quantum gravity on asymptorically $AdS_3$ spacetimes with appropriate boundary conditions is described by a 2D CFT which lives on the boundary, and total central charge of the CFT was computed there.
In these applications, the 3-dim topologically massive gravity theory was reformulated as the sum of two Chern-Simons action with complex conjugate gauge fields and complex conjugate coupling constants.~\cite{Jackiw}

Then, as mentioned in above peper, 2+1 dim quantum Hall effect is related to 2+1 dim noncommutative Chern-Simons.  Moreover 3-dim gravity can be described by Chern-Simons term.  So, it is natural question whether the 3-dim gravity theory can be expressed by the quantum Hall effect.  Then the Chern-Simons term describing the 3-dim gravity was extended to noncommutative one, and equivalence between the 3-dim Einstein gravity with negative cosmological constant and the exciton model established in our paper~\cite{our work} was pointed out in this paper.

The first part of this paper is a review of the 3-dim gravity described with Chern-Simons term~\cite{Jackiw}, in section 2, and the equivalence between quantum Hall effect and non-commutative Chern-Simons theory, in section 3.  After a review of Susskind's matrix formulation of quantum Hall effect, our exciton formulation is introduced in section 4.  In section 5, mapping between 3-dim gravity and quantum Hall effect was studied.  The last section is devoted to discussion.  

%--------------------------------
%
%         3-dim gravity
%
%--------------------------------
\section{3-dim gravity and complex Chern-Simons terms}
This section is a review of mainly the paper about 3-dim gravity described by Chern-Simons term~\cite{Jackiw}.

The action of topologically massive 3-dimensional gravity with a negative cosmological constant determining an AdS geometry is
%---
\begin{eqnarray}
 I = \int d^3 x \sqrt{g}(R + \frac{2}{l^2}) - \frac{1}{\mu}CS(\Gamma)
 \label{3d-gravity}
\end{eqnarray}
%---
where, $l$ is the radius of the curvature related to the cosmological constant by $-\Lambda = l^{-2}$. Following the conventions~\cite{chiralgravity}
$R = g^{\beta\gamma}\partial_\alpha\Gamma^\alpha_{\gamma\beta} + ...$,
%---
\begin{eqnarray}
 CS(\Gamma) = \int d^3 x\epsilon^{\alpha\beta\gamma}
           \Bigl{(}
             \frac{1}{2}\Gamma^\mu_{\nu\alpha}\partial_\beta\Gamma^\nu_{\mu\gamma}
           + \frac{1}{3}\Gamma^\mu_{\nu\alpha}\Gamma^\nu_{\sigma\beta}
                                              \Gamma^\sigma_{\mu\gamma}
           \Bigr{)}
\end{eqnarray}
%---
Using the relation between Christoffel and spin connections $\omega^{ab}_\mu$
%---
\begin{eqnarray}
 \Gamma^\alpha_{\mu\nu} = E^\alpha_a (\partial_\mu e^a_\nu
                         + \omega^a_{b\mu} e^b_\nu)
\end{eqnarray}
%---
and the 3-dimensional identity
%---
\begin{eqnarray}
 \epsilon_{abc}\epsilon^{\alpha\beta\gamma}e^c_\gamma
  = det(e)( E^\alpha_a E^\beta_b - E^\alpha_b E^\beta_a )
\end{eqnarray}
%---
 , the action (\ref{3d-gravity}) can be presented in terms of Dreibein $e^a_\mu$ in first order form as
%---
\begin{eqnarray}
 I = \int d^3 x \frac{1}{2}\epsilon_{abc}\epsilon^{\alpha\beta\gamma}e^c_\gamma
     \Bigl{(}
       {R^{ab}}_{\alpha\beta} + \frac{2}{3l^2}e^a_\alpha e^b_\beta
     \Bigr{)}
     - \frac{1}{\mu}CS(\omega)
\end{eqnarray}
%---
where,
%---
\begin{eqnarray}
 {R^{ab}}_{\alpha\beta} 
   = e^a_\mu e^b_\beta {R^{\mu\nu}}_{\alpha\beta}
   = \partial_\alpha{\omega^{ab}}_{\beta}
     - \partial_\beta{\omega^{ab}}_{\alpha}
     + {\omega^{a}}_{c\alpha}{\omega^{cb}}_{\beta}
     - {\omega^{a}}_{c\beta}{\omega^{cb}}_{\alpha}
\end{eqnarray}
%---
and,
%---
\begin{eqnarray}
 CS(\omega) = \int d^3 x \epsilon^{\alpha\beta\gamma}e^a_\alpha
              \Bigl(
               \frac{1}{2}{\omega^a}_{b\alpha}\partial_\beta{\omega^b}_{a\gamma}
               +\frac{1}{3}{\omega^a}_{b\alpha}{\omega^b}_{c\beta}
                                                      {\omega^c}_{a\gamma}
              \Bigr).
\end{eqnarray}
%---
The action $I$ is expressed in terms of the dual of the spin connection, $a_{\mu c}$,
%---
\begin{eqnarray}
 {\omega^{ab}}_\mu = -{\epsilon^{abc}}a_{\mu c}
\end{eqnarray}
%---
%---
\begin{eqnarray}
 I = \int d^3 x \epsilon^{\alpha\beta\gamma}e^a_\alpha
    \Bigl[
      -f_{\beta\gamma |a}
      +\frac{1}{3l^2}\epsilon_{abc}e^b_\beta e^c_\gamma   
    \Bigr]
    +\frac{1}{\mu}CS(a)
\end{eqnarray}
%---  
where
%---
\begin{eqnarray}
 -f_{\beta\gamma |a} = \partial_\beta a_{\gamma a}
                       -\partial_\gamma a_{\beta a}
                       +\epsilon_{abc}a^b_\beta a^c_\gamma
\end{eqnarray}
%---
%---
\begin{eqnarray}
 CS(a) = \int d^3 x \epsilon^{\alpha\beta\gamma}
         \Bigl(
          a^a_\alpha\partial_\beta a_{\gamma a}
         +\frac{1}{3}\epsilon_{abc}a^a_\alpha a^b_\beta a^c_\gamma
         \Bigr).
\end{eqnarray}
Now, introduce new quantities $\cal{A}$ and $\tilde{\cal{A}}$, which relate connection with Dreibein using arbitrary quantities m and n (In a different context and with real $m = -n$, was considered by Witten~\cite{another Witten's}),
%---
\begin{eqnarray}
 {\cal{A}}^a_\alpha = a^a_\alpha - e^a_\alpha /n, \quad
 \tilde{{\cal{A}}^a_\alpha} = a^a_\alpha - e^a_\alpha /m,\quad  
 m \neq n.
 \label{gauge-like}
\end{eqnarray}
%---
and, choose m and n to satisfy as below to remove crossterm of $\cal{A}$ and $\tilde{\cal{A}}$. 
%---
\begin{eqnarray}
 m + n = \frac{1}{\mu} \\
m \cdot n = l^2
\end{eqnarray}
%---
These equations means that $m+n$ and $m\cdot n$ must be real and positive,  
Then, the action is expressed with CS terms,
%---
\begin{eqnarray}
 I = \frac{n^2}{n-m} CS({\cal{A}}) + \frac{m^2}{m-n} CS(\tilde{\cal{A}}).
 \label{action_CS1}
\end{eqnarray}
%---
Note that m and n have a connection between $\mu$, $l$
%---
\begin{eqnarray}
 \frac{1}{\mu l}= \sqrt{\frac{m}{n}} + \sqrt{\frac{n}{m}}\geq 2
\end{eqnarray}
%---
and, if $\mu l > 1/2$, m and n must be complex, with $m^* = n$, gives that $A^\dagger =\tilde{A}$.
To include quantum theory in the topologically massive 3-dim gravity theory described as (\ref{3d-gravity}), the action is defined only for $\mu l = 1$~\cite{chiralgravity}.
These parameterization gives
%---
\begin{eqnarray}
 I = \frac{l}{2}
     \Bigl(
      (1-\frac{i}{\sqrt{3}})CS({\cal{A}}) 
     +(1+\frac{i}{\sqrt{3}})CS({\cal{A}}^\dagger) 
     \Bigr).
 \label{quantum_TMG}
\end{eqnarray}
%---
Thus, the 3-dim gravity theory can be expressed with two Chern-Simons terms, $CS({\cal{A}})$ and $CS({\cal{A}}^\dagger)$, which are disconnected from each other only if the fundamental dynamical variables are taken to be ${\cal{A}}$ and $\tilde{\cal{A}}$.

%---

%--------------------------------
%
%         NCCS and QHE
%
%--------------------------------
\section{non-Commutative Chern-Simons and Quantum Hall Effect}
Now let us review the equivalence of non-commutative Chern-Simons (NCCS) and quantum Hall effect~\cite{Susskind} briefly.

Quantum Hall effect (QHE) is the phenomenon which occurs in the 1+2 dimensional  
electron fluid under a strong magnetic field B, applied from outside. 
%---
\begin{eqnarray}
  L = \frac{eB\rho_0}{2}\epsilon_{ab}\int d^2 y
  [(\dot{x}_a - \frac{1}{2\pi\rho_0}\{x_a,A_0\})x_b
    + \frac{\epsilon_{ab}}{2\pi\rho_0}A_0]
 \label{Susskind4-7}
\end{eqnarray}
%---

The coordinates $y_1,y_2$ label the material points of the fluid, which is called the Lagrangian description.  On the other hand, $x_1,x_2$ are called the Eulerian description of which expresses the fluid properties.  $\rho_0$ is the number of particles per unit area in y space.  $A_0$ is time component of gauge field introduced for the system to be supplemented with the constraint,
%---
\begin{eqnarray}
 \frac{1}{2}\epsilon_{ij}\epsilon_{ab}
     \frac{\partial x_b}{\partial y_j}
     \frac{\partial x_a}{\partial y_i} = 1
\end{eqnarray}
%---
which means absence of vortices.  (i.e. the fluid is imcompressive.)
If one would like to introduce quasiparticle, the constraint equation, using analog of a vortex i.e. $\delta$ function, should be
%---
\begin{eqnarray}
 \frac{1}{2}\epsilon_{ij}\epsilon_{ab}
   \frac{\partial x_b}{\partial y_j}
   \frac{\partial x_a}{\partial y_i} - 1 = q\delta^2(y).
   \label{quasiparticle}
\end{eqnarray}
%---
(\ref{quasiparticle}) has the solution
%---
\begin{eqnarray}
 x_i = y_i \sqrt{1 + \frac{q}{\pi |y|^2}}
 \label{sol_QP}
\end{eqnarray}
%---
Where the $\delta$ function is
%---
\begin{eqnarray}
  \nabla\times A = 2\pi\rho_0 q\delta^2(y)
\end{eqnarray}
%---
It is convenient to introduce non-commutativity parameter
%---
\begin{eqnarray}
  \theta = \frac{1}{2\pi\rho_0}.
\end{eqnarray}
%---
And, the constraint (\ref{quasiparticle}) is equivalent to introduce a vector field A as small deviations from time independent equilibrium solution of the equations of motion, $x_i = y_i$:
%---
\begin{eqnarray}
  x_i = y_i + \epsilon_{ij}\frac{A_j}{2\pi\rho_0}
  \label{deviation}
\end{eqnarray}
%---
Then, one can find easily that the solution (\ref{sol_QP}) is identical to (\ref{deviation}), so the Lagrangian (\ref{Susskind4-7}) can be expanded into the model to express quasi-particles, using (\ref{deviation}):
%---
\begin{eqnarray}
  L^\prime = \frac{1}{4\pi\nu}\epsilon_{\mu\nu\rho}
             \Bigl[
             \frac{\partial A_\mu}{\partial y_\rho}
            -\frac{\theta}{3}\{ A_\mu, A_\rho \}
              \Bigr]A_\nu
  \label{APD}
\end{eqnarray}
%---
This is the Lagrangian of the symplectic Chern-Simons theory, and equivalent to the general form of a system on a coadjoint orbit of the group of the area preserving diffeomorphisms of some 2-dim surface~\cite{APD}.

On the other hand, one can derivate (\ref{APD}) from the NCCS with spatial non-commutativity, expanding star product to first order in $\theta$.
%---
\begin{eqnarray}
  L_{NC} = \frac{1}{4\pi\nu}\epsilon_{\mu\nu\rho}
           \Bigl(
             \hat{A}_\mu * \partial_\nu\hat{A}_\rho
            + \frac{2i}{3}\hat{A}_\mu * \hat{A}_\nu * \hat{A}_\rho
           \Bigr)
\end{eqnarray}
%---

Now we can see the equivalence between QHE and NCCS under the APD invariance.

%--------------------------------
%
%         Matrix model and Exciton sol
%
%--------------------------------
\section{Exciton in matrix formulation}
In this section, review of the exciton model advocated by authors~\cite{our work} is given after reviewing the matrix formulation of ~\cite{Susskind}.

The APD describing the quantum Hall effect is continuous one, and can describe system in long range, but it does not  work in short range property like electrons or particle.  Then, APD should be converted to discontinuous one.  One of the way is to extend the model to the matrix formulation proposed by Susskind~\cite{Susskind}.  Replacing parameter of the system with matrix, one can get indefiniteness of position.  That is, each particle obtains area to spread.  To be specific, it is to assume a classical configuration space of K electrons and replace it by a space of two $K\times K$ hermitian matrices.  Then, the diagonal elements of the matrix corresponds to coordinates of each particle.

Next, replace the Poisson bracket to classical commutator, i.e.
%---
\begin{eqnarray}
 \{ x_a(y),x_b(y)\} \rightarrow \frac{1}{i\theta}[x_a,x_b]
\end{eqnarray}
%---
Then, the action of quantum Hall system (\ref{Susskind4-7}) is generalized to matrix theory:
%---
\begin{eqnarray}
 L^\prime = \frac{eB}{2}\epsilon_{ab}Tr(\dot{x}_a - i[x_a,\hat{A}_0]_m)x_b
            + eB\theta\hat{A}_0
 \label{Susskind7-7}
\end{eqnarray}
%---
The subscript m means that the commutator is evaluated in the classical matrix space and not in the Hilbert space of quantum mechanics, in other word, the commutator originate from the Poisson bracket. 

The equation of constraint is obtained by varying this action with respect to $\hat{A}_0$, and one can easily find
%---
\begin{eqnarray}
 [x_a,x_b]_m = i\theta_{ab}.
\end{eqnarray}
%---
As is well known that above equation can be only solved with infinite matrices, so the number of electrons K must be chosen infinite.

Now, choose two matrices to satisfy
%---
\begin{eqnarray}
 [y_a,y_b]_m = i\theta_{ab} = i\theta\epsilon_{ab}
 \label{Susskind7-8}
\end{eqnarray}
%---
or, it can be represented as
%---
\begin{eqnarray}
 y_2 = -i\theta\frac{\partial}{\partial y_1}.
\end{eqnarray}
%---
Next define the matrices $\hat{A}_a$
%---
\begin{eqnarray}
 x_a = y_a + \epsilon_{ab}\theta\hat{A}_b
 \label{Susskind7-10}
\end{eqnarray}
%---
Then, one can easily see that the matrix model described as (\ref{Susskind7-7}) is identical to the Chern Simons Lagrangian (\ref{APD}) by substituting (\ref{Susskind7-10}) for (\ref{Susskind7-7}).
Now, introduce basis vectors $|m)$ where $m$ runs over the positive integers and zero.  And introduce operator $a$ and $a^\dagger$
%---
\begin{eqnarray}
 a^\dagger |n) = \sqrt{n+1} |n+1)
 \nonumber \\
 a |n) = \sqrt{n} |n-1)
\end{eqnarray}
%---
The solution which satisfies (\ref{Susskind7-8}) is expressed in terms of Fock space matrices;
%---
\begin{eqnarray}
 y_1 = \sqrt{\frac{\theta}{2}}(a + a^\dagger) 
 \nonumber \\
 y_2 = \sqrt{\frac{\theta}{2}}i(a - a^\dagger) 
\end{eqnarray}
%---
Next, consider system including quasi-particles or quasi-holes.
A quasi-particle is made when the fluid is pushed out from the center of the vortex, by the amount that proportion to the filling fraction.  And these pushed fluid is put around the hole, constructing surplus area of electrons, which is so called quasi-particle.  On the contrary, the deficit area of the fluid is quasi-hole.

And the delta function, which is analog of vortices is expressed as the projection operator onto the vector $|0)$.
Then, as was mentioned in introduction, the constraint equation describing the system of quasi-particles is given as (\ref{constraint-qe}).
If one want the quasi-holes case, one should replace $\nu$ to $-\nu$ at the second term;
%---
\begin{eqnarray}
 \theta^{-1}[x_1,x_2]_m = i - i\nu |0)(0|.
\end{eqnarray}
%---
Now, introduce two new matrices $b$,$b^\dagger$ as
%---
\begin{eqnarray}
 b^\dagger |n) = \sqrt{n+1+\nu}|n+1)\qquad\qquad\quad
 \nonumber \\
 b|n) = \sqrt{n+\nu}|n-1)\qquad for\quad n\neq 0
 \nonumber \\
 b|0) = 0 \qquad\qquad\qquad\qquad\qquad\qquad\quad
\end{eqnarray}
%---
and one can find the solution to satisfy the constraint (\ref{constraint-qe});
%---
\begin{eqnarray}
 x_1 = \sqrt{\frac{\theta}{2}}(b + b^\dagger)
 \nonumber \\
 x_2 = \sqrt{\frac{\theta}{2}}i(b - b^\dagger).
\end{eqnarray}
%---
However these model describe the case including only quasi-particles or quasi-holes, and it should be including both of them if one need more natural description.

Now, let's extend above model to express exciton.~\cite{our work}

First, introduce two independent matrices for electrons and holes.  The action of the exciton model is defined as the system of electrons and holes which are moving under the Lorentz force caused by external magnetic field each other;
%---
\begin{eqnarray}
 L_{exc} = \int dt \frac{eB}{2}\epsilon_{ab}
  \bigr [
   Tr \{ (\dot{x_e^a} + i[x_e^a,\hat{A}_0])x_e^b \}
  \nonumber \\ 
  \qquad\qquad
   - Tr \{ (\dot{x_h^a} + i[x_h^a,\hat{A}_0])x_h^b \}
   -2\theta Tr \{\hat{A}_0 \}
   \bigl ].
   \label{exciton}
\end{eqnarray}
%---
 The constraint equation is derived easily from this,
%---
\begin{eqnarray}
 [x_e^1,x_e^2]-[x_h^1,x_h^2] = i\theta\hat{1}.
 \label{exciton constraint}
\end{eqnarray}
%---
This constraint means that the area occupied by electrons contributes positively to the constraint, while that by holes does negatively, so that the minimum value of the difference of two phase space areas is to be $2\pi\theta$.
  One can see that this constraint can be converted to phase space description  by multiplying the right side by $i\theta$ and replacing $Tr$ with the phase space integral;
%---
\begin{eqnarray}
 \frac{\partial(x_e^1,x_e^2)}{\partial(y^1,y^2)}
   -\frac{\partial(x_h^1,x_h^2)}{\partial(y^1,y^2)}=1
\end{eqnarray}
%---
This shows that the fluid is incompressible. 
If the surplus of area occurs at the quasi-particle's excitation position and the deficit of the same area occurs at the quasi-hole's excitation position, it says that classical exciton solution has gained.  

Then, we begin by labeling basis vectors $|m>$ where $m$ runs over positive integers and zero.  Next we introduce two kind of lowering operators in the matrix space, $b$ and $d$ as follows:
%---
\begin{eqnarray}
 b^\dagger|n> = \sqrt{n+1+\nu}|n+1> \qquad\qquad\quad
 \nonumber \\
 b|n> = \sqrt{n+\nu}|n-1>\qquad for\quad n\neq 0
 \nonumber \\
 b|0> = 0 \qquad\qquad\qquad\qquad\qquad\qquad\qquad
\end{eqnarray}
%---
and
\begin{eqnarray}
 d^\dagger|n> = \sqrt{n+1-\nu}|n+1> \qquad\qquad\quad
 \nonumber \\
 d|n> = \sqrt{n-\nu}|n-1>\qquad for\quad n\neq 0
 \nonumber \\
 d|0> = 0. \qquad\qquad\qquad\qquad\qquad\qquad\qquad
\end{eqnarray} 
Then, the classical exciton solution $(z_e,z_h)$ in the complex notation is
%---
\begin{eqnarray}
 z_e = x_e^1+ix_e^2=\sqrt{2\theta}b
 \nonumber \\
 z_h = x_h^1+ix_h^2=\sqrt{2\theta}d^\dagger .
\end{eqnarray}
%---
It is also the solution that is added the center of mass coordinates of the quasi-electron and quasi-hole, $z_e(t)$ and $z_h(t)$ respectively to above solution, that is;
%---
\begin{eqnarray}
 z_e = z_e(t)\hat{1} + \sqrt{2\theta} b \qquad\qquad\qquad\qquad\qquad\qquad\qquad\qquad
 \nonumber \\
   = z_e(t)\sum^{\infty}_{n=0}|n><n|
     + \sqrt{2\theta}\sum^{\infty}_{n=1}\sqrt{n+\nu}|n-1><n|
 \\ 
 z_h = z_h(t)\hat{1} + \sqrt{2\theta} d^\dagger
 \qquad\qquad\qquad\qquad\qquad\qquad\qquad\qquad
 \nonumber \\
   = z_h(t)\sum^{\infty}_{n=0}|n><n|
     + \sqrt{2\theta}\sum^{\infty}_{n=1}\sqrt{n-\nu}|n><n-1|.
\end{eqnarray}
%---

%--------------------------------
%
%         3dim gravity and NCCS
%
%--------------------------------
\section{3-dim gravity described by NCCS}
In this section, I'll extend the 3-dim gravity model to non-commutative one,
 and show the equivalence to the exciton model described in matrix-like model~\cite{our work}.

The action (\ref{quantum_TMG}) expressed in the CS term in the previous section has two gauge fields, i.e. ${\cal{A}}, {\cal{A}}^\dagger$.  Let us start from this action and take the product of the fields to the star product.
Then, the action of the non-commutative 3-dim gravity model takes the form\footnote{In \cite{Banados:2001xw}, the connection between gauge fields and gravitational variables such as spin connection and triad was established so that the noncommutative Einstein equations and their corresponding action can be obtained.  There are another related references as \cite{related_1}, \cite{related_2}.}\,
%---
\begin{eqnarray}
  I_{NC} = \frac{l}{2}
     \Bigl(
      (1-\frac{i}{\sqrt{3}})CS_{NC}({\cal{A}}) 
     +(1+\frac{i}{\sqrt{3}})CS_{NC}({\cal{A}}^\dagger) 
     \Bigr).
 \label{NC_TMG}
\end{eqnarray}
%---
Here, $CS_{NC}({\cal{A}})$ is
%---
\begin{eqnarray}
  CS_{NC}({\cal{A}}) = \int d^3 x\epsilon^{\alpha\beta\gamma}
     \Bigl(
      {\cal{A}}_a^\alpha * (\partial_\beta {\cal{A}}_{\gamma a}) 
     +\frac{1}{3}\epsilon_{abc}
      {\cal{A}}^a_\alpha *{\cal{A}}^b_\beta * {\cal{A}}^c_\gamma
     \Bigr).
 \label{NCCS_TMG}
\end{eqnarray}
%---
 Expanding star-product to first order in $\theta$, which gives
%---
\begin{eqnarray}
I_{NC}=\frac{l}{2}
       \Bigl[
         (1-\frac{i}{\sqrt{3}})\int d^3 x\epsilon^{\alpha\beta\gamma}
           [(\partial_\gamma {\cal{A}}^b_\alpha){\cal{A}}_{\beta b}
             -\frac{\theta}{3}\epsilon_{abc}
              \{ {\cal{A}}^a_\alpha,{\cal{A}}^c_\gamma \} {\cal{A}}^b_\beta]
              \\
              \nonumber
               \label{NC-3dimG}
     + (1+\frac{i}{\sqrt{3}})\int d^3x \epsilon^{\alpha\beta\gamma}
         [(\partial_\gamma {\cal{A}}^{\dagger b}_\alpha)
           {\cal{A}}^\dagger_{\beta b}
            -\frac{\theta}{3}\epsilon_{abc}
             \{
              {\cal{A}}^{\dagger a}_\alpha, 
              {\cal{A}}^{\dagger c}_\gamma
             \} {\cal{A}}^{\dagger b}_\beta ]            
       \Bigr ]
\end{eqnarray}
%---
  Now, remember that the action (\ref{APD}) is equals to the action (\ref{Susskind4-7}), describing 2+1 dimensional quantum Hall system.  Then, one can easily show that (\ref{NC-3dimG}) would be like below, using C-S gauge field $A_0$ as Lagrange constant,
%---
\begin{eqnarray}
 I_{NC}=\epsilon^{\alpha\beta\gamma}\int d^3 x
        \Bigl [
         (1-\frac{i}{\sqrt{3}})
           [
           \dot{\cal{A}}^a_\alpha 
            - \frac{1}{2\pi \rho_0}
              \{ {\cal{A}}^a_\alpha, A_0 \} {\cal{A}}^b_\alpha
           ]
           \\
           \nonumber
         + (1+ \frac{i}{\sqrt{3}})
           [
             \dot{\cal{A}}^{\dagger a}_\alpha 
              - \frac{1}{2\pi \rho_0}
              \{ {\cal{A}}^{\dagger a}_\alpha, A_0 \} {\cal{A}}^{\dagger b}_\alpha
           ]
              \\
              \nonumber
              - \frac{2\epsilon_{ab}}{2\pi\rho_0}
        \Bigr ].
        \label{3dimG-exciton}
\end{eqnarray}
%---
While, the Lagrangian of the matrix-like exciton model (\ref{exciton}) is rewrite in words of the noncommutative Chern-Simons theory as below.
%---
\begin{eqnarray}
 L = \frac{eB\rho_0}{2}\epsilon_{ab}\int d^2 y
    \Bigl [
    ( \dot{x}_e^a - \frac{1}{2\pi\rho_0}\{x_e^a,A_0\} )x_e^b
   - ( \dot{x}_h^a - \frac{1}{2\pi\rho_0}\{x_h^a,A_0\} )x_h^b
   + \frac{1}{2\pi\rho_0}\epsilon_{ab}A_0
   \Bigr ]
   \nonumber
\end{eqnarray}
%---
If one interprets gauge fields ${\cal{A}}, {\cal{A}}^\dagger$ as position of electron or holes i.e. $x_e^a, x_h^a$ respectively, the action (\ref{3dimG-exciton}) would be equivalent to that of the exciton model~\cite{our work}.
Note that the Latin indices $a,b,...$, depending on the geometry of back ground space (i.e. $a,b,...=1,2,3$ in Euclidean space, or $a,b,...=0,1,2$ in Minkowski space) means label of particles that are filling 2-dim charged fluid in the quantum Hall effect model~\cite{Susskind}.  They are raised and lowered with $\eta_{ab}$ and $\eta^{ab}$. On the other hand,  Greek indices $\mu, \nu,...$ are raised and lowered with metric $g_{\mu\nu}$ or its inverse $g^{\mu\nu}$, representing indices of spacetime.

 There is one thing to notice. In the quantum Hall system~\cite{Susskind} or the exciton system~\cite{our work}, the Latin indice $a,b,...$ run $1,2$, however in this extended 3-dim gravity system, they run $1,2,3$.
 This suggests that the charged fluid would spread to 3-dim space.
%--------------------------------
%
%         conclusion
%
%-------------------------------- 
\section{Conclusion}
In this paper I have reviewed of several papers, and I have studied to show that the three dimensional Einstein gravity is related to the quantum Hall effect.

The action of topologically massive 3-dim gravity with a negative cosmological constant is described by using only Chern-Simons term~\cite{Jackiw}, while non-commutative Chern-Simons action is equivalent to that of the quantum Hall effect through the symmetry under Area Preserving Diffeomorphisms~\cite{Susskind}.  I have reviewed the derivations given in those papers.  The two independent Chern-Simons terms in ~\cite{Jackiw} seems to concern with quantum Hall effect which has two independent degree of freedom.  Then I reviewed the matrix formulated quantum Hall effect given in ~\cite{Susskind} and our exciton model, expansion of the former model into having two component matrices.

Finally in conclusion, I have examined 3-dim gravity described in the Chern-Simons term by expanding to non-commutative model, and pointed out the equivalence between the 3-dim gravity theory and our exciton model.

%--------------------------------
%
%         acknowledgments
%
%--------------------------------
\section*{Acknowledgment}
The author would like to thank Akio Sugamoto for discussions. I also give my thanks to fellows at Ochanomizu U. for giving warm reception for stray OB. And I also give my thanks to Nicol{\'a}s Grandi for commenting on my peper and introducing related references~\cite{Banados:2001xw}~\cite{related_1}~\cite{related_2}. 
%%%---

\end{document}